\documentclass[prl,twocolumn,showpacs,aps,superscriptaddress,floatfix]{revtex4}
\usepackage{amsmath}
\usepackage{amssymb}
\usepackage{bm}
\usepackage{graphicx}
\usepackage{color}
\usepackage{hyperref}
\usepackage{verbatim}

\begin{document}

\title{Externally controlled band gap in twisted bilayer graphene}

\author{A.O. Sboychakov}
\affiliation{Center for Emergent Matter Science, RIKEN, Wako-shi, Saitama,
351-0198, Japan}
\affiliation{Institute for Theoretical and Applied Electrodynamics, Russian
Academy of Sciences, Moscow, 125412 Russia}

\author{A.V. Rozhkov}
\affiliation{Center for Emergent Matter Science, RIKEN, Wako-shi, Saitama,
351-0198, Japan}
\affiliation{Institute for Theoretical and Applied Electrodynamics, Russian Academy of Sciences, Moscow, 125412 Russia}
\affiliation{Moscow Institute for Physics and Technology (State
University), Dolgoprudnyi, 141700 Russia}

\author{A.L. Rakhmanov}
\affiliation{Center for Emergent Matter Science, RIKEN, Wako-shi, Saitama,
351-0198, Japan}
\affiliation{Moscow Institute for Physics and Technology (State
University), Dolgoprudnyi, 141700 Russia}
\affiliation{Institute for Theoretical and Applied Electrodynamics, Russian Academy of Sciences, Moscow, 125412 Russia}
\affiliation{Dukhov Research Institute of Automatics, Moscow, 127055 Russia}

\author{Franco Nori}
\affiliation{Center for Emergent Matter Science, RIKEN, Wako-shi, Saitama,
351-0198, Japan}
\affiliation{Department of Physics, University of Michigan, Ann Arbor, MI
48109-1040, USA}

\begin{abstract}
We theoretically study the effects of electron-electron interaction in twisted bilayer graphene in applied transverse dc electric field. When the twist angle is not very small, the electronic spectrum of the bilayer consists of four Dirac cones inherited from each graphene layer. Applied bias voltage leads to the appearance of two hole-like and two electron-like Fermi surface sheets with perfect nesting among electron and hole components. Such a band structure is unstable with respect to exciton band gap opening due to the screened Coulomb interaction. The exciton order parameter is accompanied by the spin-density-wave order. The value of the gap depends on the twist angle. More importantly, it can be controlled by applied bias voltage which opens new directions in manufacturing of different nanoscale devices.
\end{abstract}


%
%
%
%
%
%
%
%
%
%

\date{\today}

\maketitle

\textit{Introduction}---
It is known that application of the bias voltage to AB stacked bilayer graphene opens a gap in its electronic spectrum~\cite{ourBLGreview2016}. This feature makes bilayer graphene promising for applications in electronics. Experiment shows, however, that, in many cases, the structure of the bilayer graphene samples is different from a simple AB stacking, and is characterized by a non-zero twist angle between layers~\cite{ourBLGreview2016,MeleReview,STM1,STM2,SynthesysMoire}. Twisting makes the physics of the bilayer graphene more complicated and rich. For example, it leads to the appearance of the moir\'{e} patterns -- alternating dark and bright regions seen in STM images~\cite{STM1,STM2}. For a countable set of twist angles the system has a superstructure with the period, which may coincide or be a commensurate with the moir\'{e} period~\cite{ourBLGreview2016,dSPRB}.

Rotational misorientation affects also the electronic properties of the bilayer.  Analysis in the single-particle approximation shows that one can distinguish three qualitatively different types of behavior of the spectrum at low energies. When the twist angle $\theta$ is close to commensurate value corresponding to the superstructure with considerably small size of the supercell, the spectrum has a gap at Fermi level~\cite{MelePRB1,Pankratov1,ourTBLG,ourTBLG2017}. This gap, however, is very sensitive to small variations of the twist angle, and is non-negligible only for a limited number of superstructures~\cite{ourTBLG,ourTBLG2017}. With the exception of those values, one can assume that when $\theta$ is greater  than critical value $\theta_c\sim1$--$2^{\circ}$, the electron spectrum has a linear dispersion and consists of four Dirac cones inherited from two graphene layers. For a commensurate structure, which we only study in this paper, the four Dirac points of two graphene layers are distributed between two non-equivalent corners of the superlattice Brillouin zone, forming the band structure with two doubly-degenerate Dirac cones in the corners of the superlattice Brillouin zone. The Fermi velocity of these Dirac cones, however, turns out to be $\theta$-dependent: it decreases monotonically down to zero when $\theta$ goes to critical value~\cite{dSPRL,dSPRB,PNAS}. Finally, when $\theta<\theta_c$, the spectrum at low energies is characterized by flat bands and the density of states has a peak at the Fermi level~\cite{dSPRB,ourTBLG,Morell1,NanoLettTB}.

Despite the progress in understanding the electronic properties of twisted bilayer graphene, many issues still remain unclear. First of all this concerns many-body effects, since in the majority of cases, theorists are limited to a single-electron approximation~\cite{ourBLGreview2016}. In our work, we consider the effects of electron-electron interaction for superstructures with $\theta>\theta_c$ and not too small size of the supercell, when the single-electron gap can be neglected. We also assume that the transverse electric field is applied to the bilayer. This lifts the double degeneracy of four Dirac cones of the system; two of them are shifted upwards in energy, and the other two -- downwards. As a result, two hole-like and two electron-like Fermi surface sheets appear in the system with a perfect nesting between hole-like and electron-like components. This leads to Fermi surface instability with respect to exciton band gap formation due to electron-electron interaction. We show also that the exciton order parameter is accompanied by the spin-density-wave (SDW) order. The dependence of the gap on the twist angle and on the bias voltage is analyzed. It is shown that the gap can be effectively controlled by external field, which may be useful for applications.

\textit{Geometry of twisted bilayer graphene}---
The geometry of the twisted bilayer graphene is described in details, in several papers~\cite{dSPRB,MeleReview,ourBLGreview2016}. Here we only present the basic properties necessary for further considerations. Graphene monolayer has a hexagonal crystal structure consisting of two triangular sublattices $A$ and $B$. Carbon atoms in the layer $1$ are located in positions $\mathbf{r}_{\mathbf{n}}^{1A}=\mathbf{r}_{\mathbf{n}}^{1}\equiv n\mathbf{a}_1+m\mathbf{a}_2$ and $\mathbf{r}_{\mathbf{n}}^{1B}=\mathbf{r}_{\mathbf{n}}^{1}+\bm{\delta}$ with $\mathbf{n}=(n,\,m)$ ($n$, $m$ are integers), where $\mathbf{a}_{1,2}=a(\sqrt{3},\mp1)/2$ are the graphene lattice vectors ($a=2.46$\AA) and $\bm{\delta}=a(1/\sqrt{3},0)$. The positions of atoms in layer $2$ are  $\mathbf{r}_{\mathbf{n}}^{2B}=\mathbf{r}_{\mathbf{n}}^{2}\equiv d\mathbf{e}_z+n\mathbf{a}_1'+m\mathbf{a}_2'$ and $\mathbf{r}_{\mathbf{n}}^{2A}=\mathbf{r}_{\mathbf{n}}^{2}-\bm{\delta}'$, where $\mathbf{a}_{1,2}'$, $\bm{\delta}'$  are vectors  $\mathbf{a}_{1,2}$, $\bm{\delta}$ rotated by angle $\theta$, symbol $\mathbf{e}_z$ is the unit vector along the $z$-axis, and $d=3.35$\AA\, is the interlayer distance. We also introduce vectors $\bm{\delta}^{i\alpha}\equiv\mathbf{r}_{\mathbf{n}}^{i\alpha}-\mathbf{r}_{\mathbf{n}}^{i}$, which  are independent on $\mathbf{n}$. Limiting case $\theta=0$ corresponds to the AB stacking. The superstructure exists for twist angles equal
\begin{equation}\label{theta}
\cos\theta=\frac{3m_0^2+3m_0r+r^2/2}{3m_0^2+3m_0r+r^2}\,,
\end{equation}
where $m_0$ and $r$ are co-prime positive integers. Superlattice vectors $\mathbf{R}_{1,2}$ are linear combinations of $\mathbf{a}_{1}$ and $\mathbf{a}_{2}$ with integer-valued coefficients~\cite{ourBLGreview2016}. The magnitude of these vectors is~\cite{dSPRB,MeleReview,ourBLGreview2016} $|\mathbf{R}_{1,2}|=rL/\sqrt{g}$, where $L=a/[2\sin(\theta/2)]$ is the moir\'{e} period and $g=1$ if $r\neq3n$, or $g=3$ if $r=3n$ ($n$ is integer). Thus, only superstructures with $r=1$ coincide with the moir\'{e} lattice. The number of graphene unit cells of each layer inside the supercell is $N_{sc}=(3m_0^2+3m_0r+r^2)/g$. Thus, the number of carbon atoms in the superlattice cell is equal to $4N_{sc}$.

We introduce $\mathbf{b}_{1,2}=2\pi(1/\sqrt{3},\mp1)/a$, which are the reciprocal lattice vectors of the layer $1$, and $\mathbf{b}_{1,2}'$ for layer $2$ ($\mathbf{b}_{1,2}'$ are connected to $\mathbf{b}_{1,2}$ by rotation on angle $\theta$). Vectors $\bm{{\cal G}}_{1,2}$ are the reciprocal vectors for superlattice. All these vectors are related to each other according to the following formulas: $\mathbf{b}_1'=\mathbf{b}_1-r\bm{{\cal G}}_{1}$ and $\mathbf{b}_2'=\mathbf{b}_2+r(\bm{{\cal G}}_{1}+\bm{{\cal G}}_{2})$ if $r\neq3n$, or $\mathbf{b}_1'=\mathbf{b}_1+r(\bm{{\cal G}}_{1}+2\bm{{\cal G}}_{2})/3$ and $\mathbf{b}_2'=\mathbf{b}_2-r(2\bm{{\cal G}}_{1}+\bm{{\cal G}}_{2})/3$, otherwise. Each graphene layer has two non-equivalent Dirac points located at corners of its Brillouin zone. Thus, the total number of Dirac points for the bilayer is four. The Brillouin zone of the superlattice has a shape of hexagon. It can be obtained by $N_{sc}$ times folding of the Brillouin zone of the layer $1$ or $2$. As a result of this folding, Dirac points of each layer are translated to two non-equivalent Dirac points of the superlattice, $\mathbf{K}_1$ and  $\mathbf{K}_2$, located at corners of the reduced Brillouin zone. Thus, one can say that the Dirac points $\mathbf{K}_{1,2}$ are doubly degenerate since each of them corresponds to two non-equivalent Dirac points of constituent layers. Points $\mathbf{K}_{1,2}$ can be expressed via vectors $\bm{{\cal G}}_{1,2}$ as $\mathbf{K}_1=(\bm{{\cal G}}_{1}+2\bm{{\cal G}}_{2})/3$ and $\mathbf{K}_2=(2\bm{{\cal G}}_{1}+\bm{{\cal G}}_{2})/3$.

\textit{Model Hamiltonian}---
We start from the tight-binding model for the $p_z$ electrons in twisted bilayer graphene. Electrons are assumed to interact via the screened Coulomb interaction. Formally, we write
\begin{eqnarray}
H\!\!&=&\!\!\!\sum_{{i\mathbf{n}j\mathbf{m}\atop\alpha\beta\sigma}}t(\mathbf{r}_{\mathbf{n}}^{i\alpha};\mathbf{r}_{\mathbf{m}}^{j\beta})
d^{\dag}_{\mathbf{n}i\alpha\sigma}d^{\phantom{\dag}}_{\mathbf{m}j\beta\sigma}+
\frac{V_b}{2}\sum_{\mathbf{n}}\left(n_{\mathbf{n}1}-n_{\mathbf{n}2}\right)+\nonumber\\
&&\!\!\!\frac12\!\!\sum_{{i\mathbf{n}j\mathbf{m}\atop\alpha\beta\sigma\sigma'}}\!\!d^{\dag}_{\mathbf{n}i\alpha\sigma}d^{\phantom{\dag}}_{\mathbf{n}i\alpha\sigma}
V(\mathbf{r}_{\mathbf{n}}^{i\alpha}-\mathbf{r}_{\mathbf{m}}^{j\beta})d^{\dag}_{\mathbf{m}j\beta\sigma'}d^{\phantom{\dag}}_{\mathbf{m}j\beta\sigma'}\,,\label{H}
\end{eqnarray}
where $d^{\dag}_{\mathbf{n}i\alpha\sigma}$ and $d^{\phantom{\dag}}_{\mathbf{n}i\alpha\sigma}$ are the creation and annihilation operators of the electron with spin projection $\sigma$, located at site $\mathbf{n}$ in the layer $i$($=1,2$) in the sublattice $\alpha$($=A,B$), and $n_{\mathbf{n}i}=\sum_{\alpha\sigma}d^{\dag}_{\mathbf{n}i\alpha\sigma}d^{\phantom{\dag}}_{\mathbf{n}i\alpha\sigma}$. The first term describes the intersite hopping. For intralayer hopping we consider only the nearest-neighbor term with amplitude $-t$, where $t=2.57$\,eV. The interlayer hopping amplitudes are parameterized as described in Refs.~\cite{ourTBLG,ourTBLG2017}, with the largest interlayer hopping amplitude equal to $t_0=0.4$\,eV. Second term describes the potential energy difference between layers due to the applied bias voltage $V_b$. Third term corresponds to the Coulomb interaction between electrons. The precise form of the function $V(\mathbf{r})$ will be discussed below.

Let us first consider single-particle part of the Hamiltonian~\eqref{H} (first and second terms). We proceed to the momentum representation introducing electronic operators
\begin{equation}\label{dpG}
d^{\phantom{\dag}}_{\mathbf{pG}i\alpha\sigma}=\frac{1}{\sqrt{\cal N}}\sum_{\mathbf{n}}
e^{-i(\mathbf{p}+\mathbf{G})\mathbf{r}_{\mathbf{n}}^{i}}d_{\mathbf{n}i\alpha\sigma}\,,
\end{equation}
where ${\cal N}$ is the number of graphene unit cells in the sample in one layer, momentum $\mathbf{p}$ lies in the first Brillouin zone of the superlattice, while $\mathbf{G}=n\bm{{\cal G}}_1+m\bm{{\cal G}}_2$ is the reciprocal vector of the superlattice, lying in the first Brillouin zone of the $i$th layer. The number of such vectors $\mathbf{G}$ is equal to $N_{sc}$ for each graphene layer. In this representation the single-particle part of Hamiltonian~\eqref{H} can be written as
\begin{eqnarray}
H_0\!\!&=&\!\!\!\sum_{\mathbf{p}}\!\left[\!\sum_{{\mathbf{G}_1\mathbf{G}_2\atop ij\alpha\beta\sigma}}\tilde{t}_{ij}^{\alpha\beta}(\mathbf{p}+\mathbf{G}_1;\mathbf{G}_1-\mathbf{G}_2)
d^{\dag}_{\mathbf{pG}_1i\alpha\sigma}d^{\phantom{\dag}}_{\mathbf{pG}_2j\beta\sigma}\right.\!\!+\nonumber\\
&&\!\!\left.\frac{V_b}{2}\!\!\sum_{{\mathbf{G}\alpha\sigma\atop }}\left(
d^{\dag}_{\mathbf{pG}1\alpha\sigma}d^{\phantom{\dag}}_{\mathbf{pG}1\alpha\sigma}-
d^{\dag}_{\mathbf{pG}2\alpha\sigma}d^{\phantom{\dag}}_{\mathbf{pG}2\alpha\sigma}\right)\right],\label{H0}
\end{eqnarray}
where
\begin{equation}\label{tG}
\tilde{t}_{ij}^{\alpha\beta}(\mathbf{k};\mathbf{G})=\frac{1}{N_{sc}}\mathop{{\sum}'}_{\mathbf{nm}}e^{-i\mathbf{k}(\mathbf{r}_{\mathbf{n}}^{i}-\mathbf{r}_{\mathbf{m}}^{j})}
e^{-i\mathbf{G}\mathbf{r}_{\mathbf{m}}^{j}}t(\mathbf{r}_{\mathbf{n}}^{i\alpha};\mathbf{r}_{\mathbf{m}}^{j\beta})\,.
\end{equation}
In the last formula, the summation over $\mathbf{m}$ is performed over sites inside the zeroth supercell, while summation over $\mathbf{n}$ is performed over all sites in the sample (or vice versa).

\begin{figure}[t]
\centering
\includegraphics[width=0.59\columnwidth]{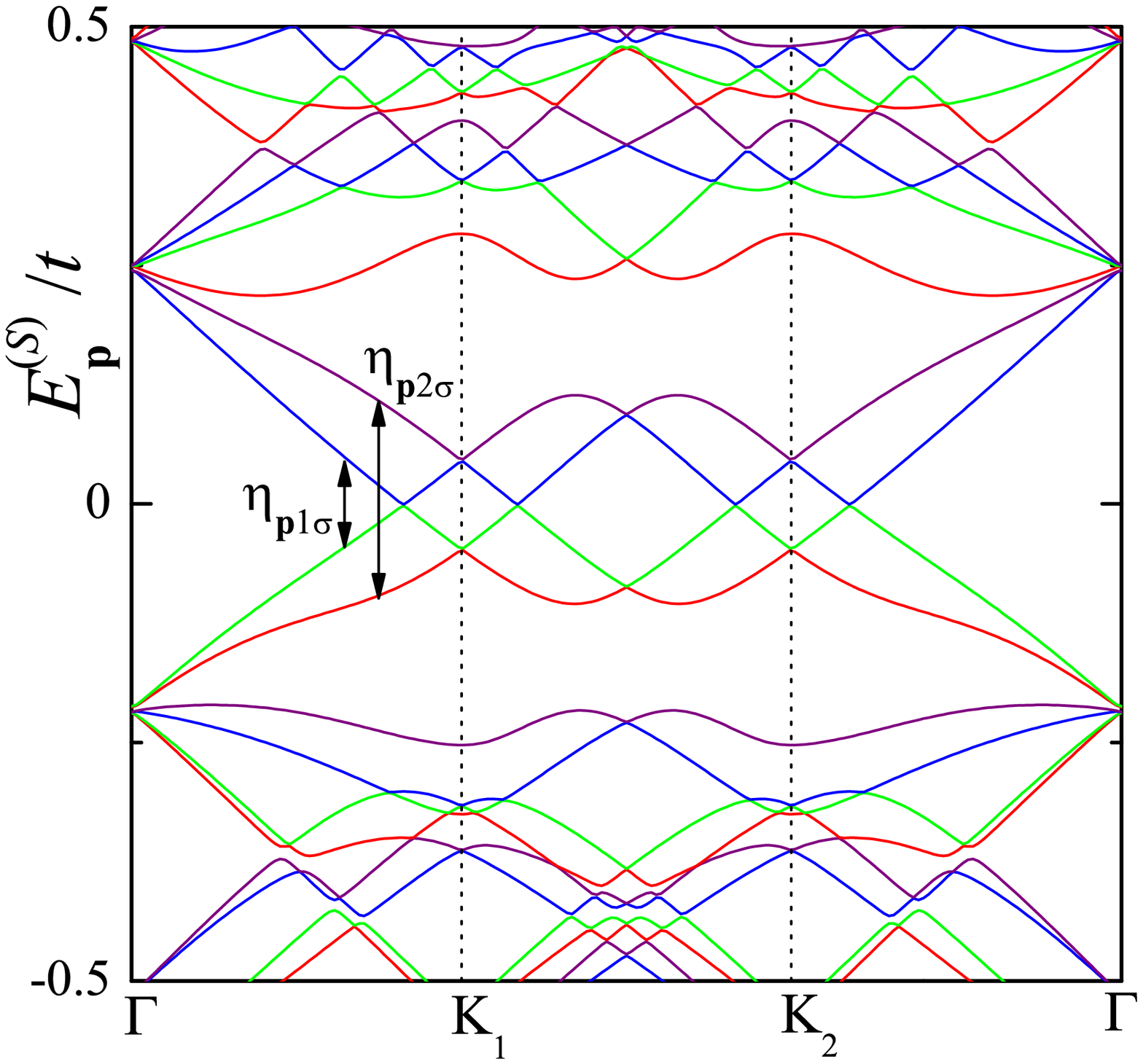}
\includegraphics[width=0.38\columnwidth]{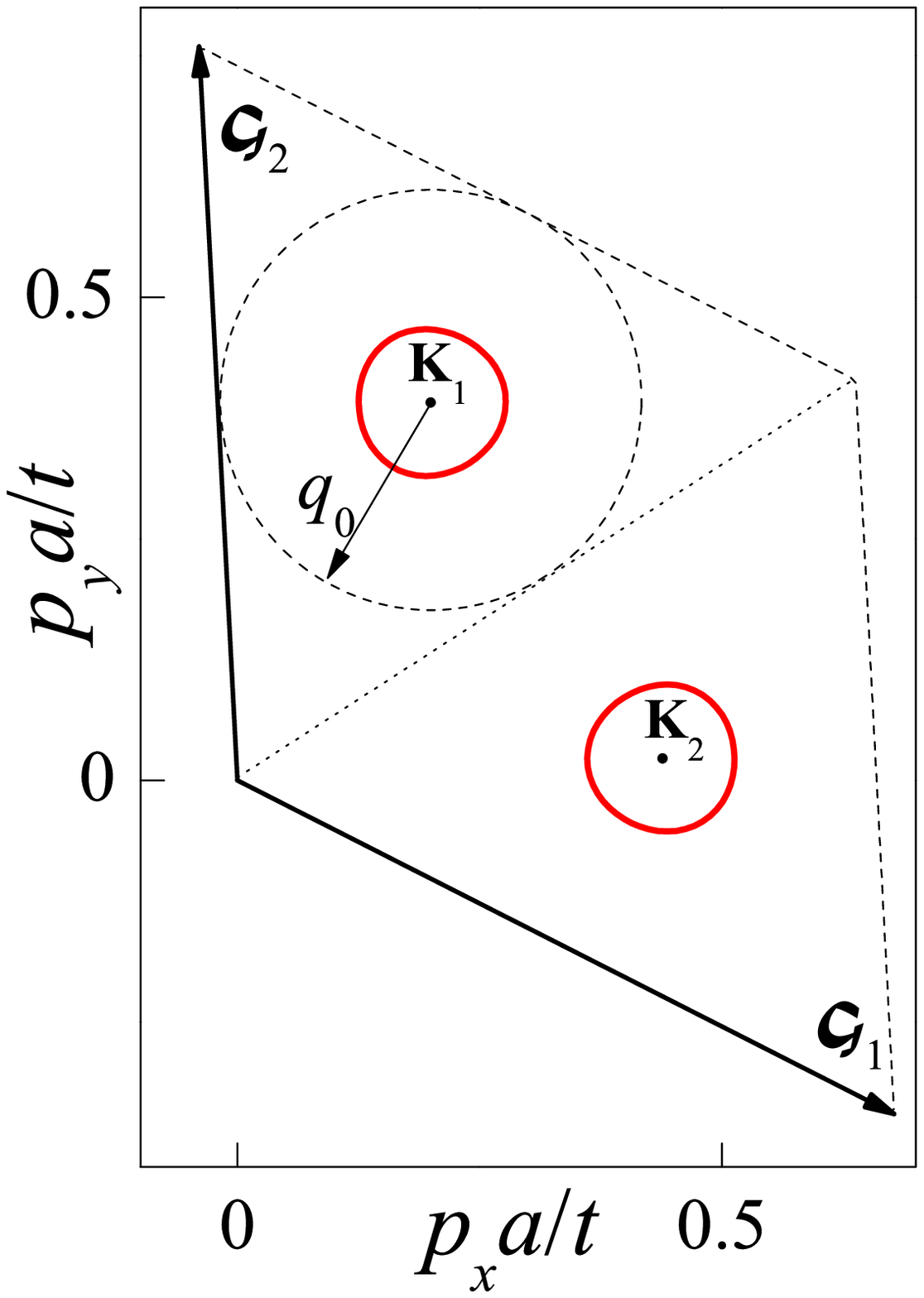}
\caption{(Left panel) Band structure calculated for the sample with $m_0=5$, $r=1$ ($\theta\cong6.009^{\circ}$); $V_b=0.15t/e$. Vertical arrows show the bands, which are coupled by the electron-electron interaction. (Right panel) Fermi surface corresponding to the band structure shown on the left. Slight trigonal warping of the Fermi surface curves is seen. Dashed circle with radius $q_0$ is used to find approximate solution to the gap equation in the limit of strong interaction (see the text).\label{FigSpec}}
\end{figure}

For a given momentum $\mathbf{p}$, Hamiltonian~\eqref{H0} can be represented in the form of $4N_{sc}\times4N_{sc}$ matrix. We diagonalize this matrix numerically, and calculate both the spectrum $E_{\mathbf{p}}^{(S)}$ ($S=1,\,2,\,\dots,\,4N_{sc}$) and eigenvectors $U^{(S)}_{\mathbf{pG}i\alpha}$. Figure~\ref{FigSpec} shows the band structure calculated for the sample with $m_0=5$, $r=1$ ($\theta\cong6.009^{\circ}$), and for bias voltage $V_b=0.15t$. Only bands within the energy window $|E|/t<0.5$ are shown. Low-energy spectrum consists of four Dirac cones located in pairs near two Dirac points of the superlattice, $\mathbf{K}_1$ and $\mathbf{K}_2$. Bias voltage shifts apexes of two of these cones to positive energies, and two -- to negative energies. The electron density corresponding to upper (lower) Dirac cones is concentrated mostly in layer 1 (layer 2), even though the interlayer hybridization tends to distribute it uniformly between the layers. When bias voltage is applied, the system acquires a Fermi surface, which consists of two closed curves (valleys) located near two Dirac points. These curves are nearly circular when the bias voltage is small enough, while trigonal warping reveals itself at larger $V_b$ (see Fig.~\ref{FigSpec}). The trigonal warping becomes more pronounced for superstructures with smaller $\theta$.

The most important feature of the Fermi surface shown in Fig.~\ref{FigSpec} is its double degeneracy; in each valley, the Fermi surface curve is created by both the electron-like band corresponding to the lower Dirac cone, and the hole-like band belonging to the upper Dirac cone. In other words, we have a situation with perfect nesting of Fermi surface. This leads to the instability of a Fermi liquid state with respect to the formation of some type of ordering due to the electron-electron interaction.

Consider now the interaction term of Hamiltonian~\eqref{H}. In terms of electron operators $d^{\phantom{\dag}}_{\mathbf{pG}i\alpha\sigma}$, Eq.~\eqref{dpG}, the last term in Eq.~\eqref{H} takes the form
\begin{eqnarray}
&&\!\!\!\!\!\!\frac{1}{2{\cal N}}\sum_{i\alpha\sigma\atop j\beta\sigma'}\sum_{\mathbf{p}_1\mathbf{p}_2\atop\mathbf{p}_1'\mathbf{p}_2'}
\sum_{\mathbf{G}_1\mathbf{G}_2\atop\mathbf{G}_1'\mathbf{G}_2'}
\sum_{\mathbf{G}}\delta_{\mathbf{G},\,\mathbf{k}_1+\mathbf{k}_2-\mathbf{k}_1'-\mathbf{k}_2'}\times\label{U}\\
&&\!\!\!\!\!\!d^{\dag}_{\mathbf{p}_1\mathbf{G}_1i\alpha\sigma}d^{\phantom{\dag}}_{\mathbf{p}_1'\mathbf{G}_1'i\alpha\sigma}
\!V_{i\alpha;j\beta}\!\left(\mathbf{k}_1-\mathbf{k}_1';\mathbf{G}\right)\!
d^{\dag}_{\mathbf{p}_2\mathbf{G}_2j\beta\sigma'}d^{\phantom{\dag}}_{\mathbf{p}_2'\mathbf{G}_2'j\beta\sigma'},\nonumber
\end{eqnarray}
where $\mathbf{k}_{1,2}=\mathbf{p}_{1,2}+\mathbf{G}_{1,2}$, $\mathbf{k}_{1,2}'=\mathbf{p}_{1,2}'+\mathbf{G}_{1,2}'$, and
\begin{equation}\label{VkG}
V_{i\alpha;j\beta}\left(\mathbf{k};\mathbf{G}\right)=\frac{1}{N_{sc}}\!\mathop{{\sum}'}_{\mathbf{nm}}
e^{-i\mathbf{k}(\mathbf{r}_{\mathbf{n}}^{i}-\mathbf{r}_{\mathbf{m}}^{j})}
e^{-i\mathbf{G}\mathbf{r}_{\mathbf{m}}^{j}}V(\mathbf{r}_{\mathbf{n}}^{i\alpha}-\mathbf{r}_{\mathbf{m}}^{j\beta})\,.
\end{equation}
As before, in the last equation the summation over $\mathbf{m}$ is performed over sites inside the zeroth supercell, while summation over $\mathbf{n}$ is performed over all sites in the sample. For intralayer ($i=j$) interaction one can separate the summation on $\mathbf{n}$ and $\mathbf{m}$ by substitution $\mathbf{n}\to\mathbf{n}+\mathbf{m}$. As a result, we obtain $V_{i\alpha;i\beta}\left(\mathbf{k};\mathbf{G}\right)=\left(\sum_{\mathbf{b}^{i}}\delta_{\mathbf{b}^i\!,\,\mathbf{G}}\right)
\sum_{\mathbf{n}}e^{-i\mathbf{k}\mathbf{r}_{\mathbf{n}}^{i}}V(\mathbf{r}_{\mathbf{n}}^{i}+\bm{\delta}^{i\alpha}-\bm{\delta}^{i\beta})$, where the first and second summations are performed over all reciprocal lattice vectors ($\mathbf{b}^{i}$) and all lattice sites of the layer $i$, correspondingly. Let us denote by $\tilde{\mathbf{k}}^{i}$ the `vector $\mathbf{k}$ modulo $\mathbf{b}^{i}$', that is, the vector lying in the first Brillouin zone of the layer $i$ and coinciding with $\mathbf{k}$ upon the translation on some reciprocal vector $\mathbf{b}^{i}$. By definition~\eqref{VkG}, we have $V_{i\alpha;i\beta}(\mathbf{k};\mathbf{G})=V_{i\alpha;i\beta}(\tilde{\mathbf{k}}^{i};\mathbf{G})$. Below we will use the continuum (low-$\mathbf{k}$) approximation for $V_{i\alpha;i\beta}\left(\mathbf{k};\mathbf{G}\right)$, when one can substitute the summation over lattice sites by the 2D integration. As a result, we obtain
\begin{equation}\label{VkGii}
V_{i\alpha;i\beta}\left(\mathbf{k};\mathbf{G}\right)=\frac{1}{{\cal V}_{c}}V_{ii}(\tilde{\mathbf{k}}^{i})
e^{i\tilde{\mathbf{k}}^{i}(\bm{\delta}^{i\alpha}-\bm{\delta}^{i\beta})}\sum_{\mathbf{b}^{i}}\delta_{\mathbf{b}^i\!,\,\mathbf{G}}\,,
\end{equation}
where ${\cal V}_c=\sqrt{3}a^2/2$ is the graphene's unit cell area, and $V_{ii}(\mathbf{k})=\int d^2rV(\mathbf{r})e^{-i\mathbf{kr}}$ is the Fourier transform of the function $V(\mathbf{r})$.

We introduce also the Fourier transform for interlayer interaction  as $V_{ij}(\mathbf{k})=\int d^2rV(d\mathbf{e}_z+\mathbf{r})e^{-i\mathbf{kr}}$ ($i\neq j$). Substituting this equation to Eq.~\eqref{VkG}, one obtains
\begin{eqnarray}
V_{i\alpha;j\beta}\left(\mathbf{k};\mathbf{G}\right)&=&\frac{1}{{\cal V}_{c}}\sum_{\mathbf{b}^i}V_{ij}\left(\mathbf{k}+\mathbf{b}^i\right)
e^{i(\mathbf{k}+\mathbf{b}^i)(\bm{\delta}^{i\alpha}-\bm{\delta}^{i\beta})}\times\nonumber\\
&&\sum_{\mathbf{b}^{j}}\delta_{\mathbf{b}^j\!,\,\mathbf{b}^i+\mathbf{G}}\,.\label{VkGij}
\end{eqnarray}
For functions $V_{ij}(\mathbf{k})$ we use expressions for the screened Coulomb potential in the form~\cite{LozovikSokolik,LozovikPLA2009}:
\begin{equation}\label{Vrpa}
V_{ij}(\mathbf{q})=\frac{v_{\mathbf{q}}}{1+\Pi_{\mathbf{q}}v_{\mathbf{q}}}\left(
\begin{array}{cc}
1&e^{-qd}\\e^{-qd}&1
\end{array}\right)\,,
\end{equation}
where bare Coulomb potential is $v_{\mathbf{q}}=2\pi e^2/\epsilon q$. The permittivity of the substrate is $\epsilon$ and $\Pi_{\mathbf{q}}\equiv-P(\omega=0,\mathbf{q})$, where $P(\omega,\mathbf{q})$ is the polarization operator of the bilayer. Since interlayer potential $V_{ij}(\mathbf{q})$ decays exponentially with $q$ and $e^{-|\mathbf{b}_{1,2}|d}\approx5\times10^{-5}$, one can take only one term in the sum over $\mathbf{b}^{i}$ in Eq.~\eqref{VkGij}, such that  $\mathbf{k}+\mathbf{b}^i=\tilde{\mathbf{k}}^i$. Below we will use the long wave length approximation for the function $\Pi_{\mathbf{q}}$. In this case, it is independent on $\mathbf{q}$ and equals to the density of states of the bilayer at the Fermi level. We calculate the latter quantity following procedure described below.

\textit{Exciton order parameter}---
To go further, we introduce new electronic operators $\psi_{\mathbf{p}S\sigma}$ defined according to the relations:
\begin{equation}
\psi_{\mathbf{p}S\sigma}=\sum_{\mathbf{G}i\alpha}U^{(S)*}_{\mathbf{p}\mathbf{G}i\alpha}d_{\mathbf{pG}i\alpha\sigma}\,,\;\;
d_{\mathbf{pG}i\alpha\sigma}=\sum_{S}U^{(S)}_{\mathbf{p}\mathbf{G}i\alpha}\psi_{\mathbf{p}S\sigma}\,.
\end{equation}
In this representation,  the single-particle part of the Hamiltonian~\eqref{H} is diagonal. The interaction term takes a form $\sum V(\mathbf{p}_1S_1\dots\mathbf{p}_4S_4)
\psi^{\dag}_{\mathbf{p}_1S_1\sigma}\psi^{\phantom{\dag}}_{\mathbf{p}_2S_2\sigma}\psi^{\dag}_{\mathbf{p}_3S_3\sigma'}\psi^{\phantom{\dag}}_{\mathbf{p}_4S_4\sigma'}$, where the summation is performed over all indices in this expression, and $V(\mathbf{p}_1S_1;\mathbf{p}_2S_2;\mathbf{p}_3S_3;\mathbf{p}_4S_4)$ is the convolution of the product of the function $V_{i\alpha;j\beta}(\mathbf{k};\mathbf{G})$ and four wave functions $U^{(S)}_{\mathbf{pG}i\alpha}$.

Let us denote the indices of two electron-like and two hole-like bands closest to the Fermi level by $S^{\mu}_{+}$ and $S^{\mu}_{-}$, correspondingly, where $\mu=1,2$. The perfect nesting of hole-like and electron-like Fermi surfaces gives rise to a dielectric order parameter driven by the Coulomb interaction. In general, this order parameter is a superposition of the expectation values of the type $\langle\psi^{\dag}_{\mathbf{p}S\sigma}\psi^{\phantom{\dag}}_{\mathbf{p}S'\sigma'}\rangle$, where $S\neq S'$. Here we assume that non-zero expectation values are only those, which couple the bands $S^{\mu}_{+}$ and $S^{\mu}_{-}$ (see Fig.~\ref{FigSpec}). Analysis shows also that corresponding order is of the SDW type. Assuming planar spin configuration, we obtain that non-zero expectation values are the following
\begin{equation}\label{eta}
\eta_{\mathbf{p}\mu\sigma}=\left\langle\psi^{\dag}_{\mathbf{p}S^{\mu}_{+}\sigma}\psi^{\phantom{\dag}}_{\mathbf{p}S^{\mu}_{-}\bar{\sigma}}\right\rangle=
\left\langle\psi^{\dag}_{\mathbf{p}S^{\mu}_{-}\bar{\sigma}}\psi^{\phantom{\dag}}_{\mathbf{p}S^{\mu}_{+}\sigma}\right\rangle^{*}\!,
\end{equation}
where $\bar{\sigma}$ means `not $\sigma$'. Note that possible charge-density-wave order, corresponding to the non-zero $\tilde{\eta}_{\mathbf{p}\mu}=\sum_{\sigma}\langle\psi^{\dag}_{\mathbf{p}S^{\mu}_{+}\sigma}\psi^{\phantom{\dag}}_{\mathbf{p}S^{\mu}_{-}\sigma}\rangle$ is energetically unfavorable in comparison to the SDW ones.

We will study the model~\eqref{H} in the mean-field approximation. This scheme involves the replacement of the product of two operators $O_1O_2\to O_1\langle O_2\rangle+\langle O_1\rangle O_2-\langle O_1\rangle\langle O_2\rangle$, where $\langle O_{1,2}\rangle\propto\eta_{\mathbf{p}\mu\sigma}$. As a result, the interaction term of the Hamiltonian becomes quadratic in $\psi^{\phantom{\dag}}_{\mathbf{p}S\sigma}$ operators. In addition, we truncate the total mean-field Hamiltonian keeping only bands $S^{\mu}_{\pm}$. Moreover, in the interaction part of this Hamiltonian we keep only terms coupling the electron band $S^{\mu}_{+}$ with the hole band $S^{\mu}_{-}$ with the same $\mu$. As a result, the effective mean-field Hamiltonian becomes
\begin{equation}\label{HMF}
H_{\text{MF}}=\sum_{\mathbf{p}\mu}\Psi^{\dag}_{\mathbf{p}\mu}\hat{H}_{\mathbf{p}\mu}\Psi^{\phantom{\dag}}_{\mathbf{p\mu}}+U_{c}\,,
\end{equation}
where we introduce the following 4-component operator $\Psi^{\phantom{\dag}}_{\mathbf{p}\mu}=(\psi^{\phantom{\dag}}_{\mathbf{p}S^{\mu}_{+}\uparrow},\psi^{\phantom{\dag}}_{\mathbf{p}S^{\mu}_{-}\uparrow},
\psi^{\phantom{\dag}}_{\mathbf{p}S^{\mu}_{+}\downarrow},\psi^{\phantom{\dag}}_{\mathbf{p}S^{\mu}_{-}\downarrow})^{\text{T}}$.
In equation~\eqref{HMF}, $\hat{H}_{\mathbf{p}\mu}$ is the $4\times4$ matrix
\begin{equation}
\hat{H}_{\mathbf{p}\mu}=\left(\begin{array}{cccc}
E^{(S^{\mu}_{+})}_{\mathbf{p}}&0&0&-\Delta_{\mathbf{p}\mu\uparrow}^{*}\\
0&E^{(S^{\mu}_{-})}_{\mathbf{p}}&-\Delta_{\mathbf{p}\mu\downarrow}&0\\
0&-\Delta_{\mathbf{p}\mu\downarrow}^{*}&E^{(S^{\mu}_{+})}_{\mathbf{p}}&0\\
-\Delta_{\mathbf{p}\mu\uparrow}&0&0&E^{(S^{\mu}_{-})}_{\mathbf{p}}
\end{array}\right),
\end{equation}
where $\Delta_{\mathbf{p}\mu\sigma}$ is the order parameter having the form
\begin{equation}\label{DeltaSigma}
\Delta_{\mathbf{p}\mu\sigma}=\frac{1}{{\cal N}}\sum_{\mathbf{q}\nu}\left[A_{\mu\nu}(\mathbf{p};\mathbf{q})\eta_{\mathbf{q}\nu\sigma}+
B_{\mu\nu}(\mathbf{p};\mathbf{q})\eta^{*}_{\mathbf{q}\nu\bar{\sigma}}\right],
\end{equation}
where  $A_{\mu\nu}(\mathbf{p};\mathbf{q})\equiv V(\mathbf{p}S^{\mu}_{-};\mathbf{q}S^{\nu}_{-};\mathbf{q}S^{\nu}_{+};\mathbf{p}S^{\mu}_{+})$ and $B_{\mu\nu}(\mathbf{p};\mathbf{q})\equiv V(\mathbf{p}S^{\mu}_{-};\mathbf{q}S^{\nu}_{+};\mathbf{q}S^{\nu}_{-};\mathbf{p}S^{\mu}_{+})$. The term  $U_{c}$ in Eq.~\eqref{HMF} is the $c$-number
\begin{eqnarray}\label{Uc}
U_{c}&=&\frac{1}{2{\cal N}}\sum_{\mathbf{pq}\atop\mu\nu\sigma}\left[A_{\mu\nu}(\mathbf{p};\mathbf{q})\eta^{*}_{\mathbf{p}\mu\sigma}\eta^{\phantom{*}}_{\mathbf{q}\nu\sigma}+\right.\nonumber\\
&&\left.B_{\mu\nu}(\mathbf{p};\mathbf{q})\eta^{*}_{\mathbf{p}\mu\sigma}\eta^{*}_{\mathbf{q}\nu\bar{\sigma}}+H.c.\right].
\end{eqnarray}
The precise form of the functions $A_{\mu\nu}(\mathbf{p};\mathbf{q})$ and $B_{\mu\nu}(\mathbf{p};\mathbf{q})$ is the following:
\begin{eqnarray}
A_{\mu\nu}(\mathbf{p};\mathbf{q})\!\!&=&\!\!\!\sum_{i\alpha\atop j\beta}\!\!\sum_{\mathbf{G}_1\mathbf{G}_2\atop\mathbf{G}_1'\mathbf{G}_2'}\!
U^{(S^{\mu}_{-})*}_{\mathbf{pG}_1i\alpha}U^{(S^{\nu}_{-})}_{\mathbf{qG}_1'i\alpha}U^{(S^{\nu}_{+})*}_{\mathbf{qG}_2j\beta}U^{(S^{\mu}_{+})}_{\mathbf{pG}_2'j\beta}\times\nonumber\\
&&\!\!\!\!\!\!\!V_{i\alpha;j\beta}(\mathbf{p}\!-\!\mathbf{q}\!+\!\mathbf{G}_1\!-\!\mathbf{G}_1';\mathbf{G}_1\!+\!\mathbf{G}_2\!-\!\mathbf{G}_1'\!-\!\mathbf{G}_2'),\nonumber\\
B_{\mu\nu}(\mathbf{p};\mathbf{q})\!\!&=&\!\!\!\sum_{i\alpha\atop j\beta}\!\!\sum_{\mathbf{G}_1\mathbf{G}_2\atop\mathbf{G}_1'\mathbf{G}_2'}\!
U^{(S^{\mu}_{-})*}_{\mathbf{pG}_1i\alpha}U^{(S^{\nu}_{+})}_{\mathbf{qG}_1'i\alpha}U^{(S^{\nu}_{-})*}_{\mathbf{qG}_2j\beta}U^{(S^{\mu}_{+})}_{\mathbf{pG}_2'j\beta}\times\nonumber\\
&&\!\!\!\!\!\!\!V_{i\alpha;j\beta}(\mathbf{p}\!-\!\mathbf{q}\!+\!\mathbf{G}_1\!-\!\mathbf{G}_1';\mathbf{G}_1\!+\!\mathbf{G}_2\!-\!\mathbf{G}_1'\!-\!\mathbf{G}_2').\nonumber\\\label{AB}
\end{eqnarray}

Minimizing the total energy at zero temperature and at half-filling, we obtain the system of equations for the order parameters:
\begin{equation}\label{DeltaPSigma}
\Delta_{\mathbf{p}\mu\sigma}\!=\!\frac{1}{2{\cal N}}\!\sum_{\mathbf{q}\nu}\!\!\left[\frac{A_{\mu\nu}(\mathbf{p};\mathbf{q})\Delta_{\mathbf{q}\nu\sigma}}{\sqrt{|\Delta_{\mathbf{q}\nu\sigma}|^2+E^2_{\mathbf{q}\nu}}}+
\frac{B_{\mu\nu}(\mathbf{p};\mathbf{q})\Delta^{*}_{\mathbf{q}\nu\bar{\sigma}}}{\sqrt{|\Delta_{\mathbf{q}\nu\bar{\sigma}}|^2+E^2_{\mathbf{q}\nu}}}\right]\!\!,
\end{equation}
where $E_{\mathbf{q}\mu}=[E^{(S^{\mu}_{+})}_{\mathbf{q}}-E^{(S^{\mu}_{-})}_{\mathbf{q}}]/2$.

For a given superstructure and bias voltage we calculate the functions $A_{\mu\nu}(\mathbf{p};\mathbf{q})$ and $B_{\mu\nu}(\mathbf{p};\mathbf{q})$ numerically. Analysis shows, that with a good accuracy the following relations take place:
\begin{eqnarray}
A_{\mu\nu}(\mathbf{p};\mathbf{q})&\approx&\delta_{\mu\nu}A(\mathbf{p};\mathbf{q})\,,\nonumber\\
B_{\mu\nu}(\mathbf{p};\mathbf{q})&\approx&\delta_{\mu\bar{\nu}}B(\mathbf{p};\mathbf{q})\,,\label{SymAB}
\end{eqnarray}
where $\bar{\nu}$ means `not $\nu$'. The deviations from these equalities do not exceed $1\%$ for any superstructures considered. Note that  in the limit of uncoupled graphene layers, $t_0=0$, Eqs.~\eqref{SymAB} become exact. This follows from the symmetry of the wavefunctions presented in the definition of the functions  $A_{\mu\nu}$ and $B_{\mu\nu}$, Eqs.~\eqref{AB}: for uncoupled layers the electrons are localized either in layer $1$ or $2$.  Further, by choosing appropriate phases of the wave functions $U^{(S^{\mu}_{\pm})}_{\mathbf{pG}i\alpha}$ one can make $A(\mathbf{p};\mathbf{q})$ and $B(\mathbf{p};\mathbf{q})$ real-valued functions. In this case, one can choose $\Delta_{\mathbf{p}\mu\sigma}$ to be real-valued functions satisfying the relations $\Delta_{\mathbf{p}\mu\uparrow}=\Delta_{\mathbf{p}\mu\downarrow}\equiv\Delta_{\mathbf{p}\mu}$. The equation for the order parameters then becomes
\begin{equation}\label{DeltaP}
\Delta_{\mathbf{p}\mu}=\frac12\int\!\frac{d^2q}{v_{BZ}}\left[\frac{A(\mathbf{p};\mathbf{q})\Delta_{\mathbf{q}\mu}}{\sqrt{\Delta_{\mathbf{q}\mu}^2+E^2_{\mathbf{q}\mu}}}+
\frac{B(\mathbf{p};\mathbf{q})\Delta_{\mathbf{q}\bar{\mu}}}{\sqrt{\Delta_{\mathbf{q}\bar{\mu}}^2+E^2_{\mathbf{q}\bar{\mu}}}}
\right],
\end{equation}
where $v_{BZ}=8\pi^2/(a^2\sqrt{3})$ is the area of the graphene's Brillouin zone, while the integration is performed over the Brillouin zone of the superlattice. In numerical calculations we take $A(\mathbf{p};\mathbf{q})=[A_{11}(\mathbf{p};\mathbf{q})+A_{22}(\mathbf{p};\mathbf{q})]/2$ and $B(\mathbf{p};\mathbf{q})=[B_{12}(\mathbf{p};\mathbf{q})+B_{21}(\mathbf{p};\mathbf{q})]/2$.

\textit{Approximate solution, weak interaction limit}---
The main interest is the value of the function $\Delta_{\mathbf{p}1}$ at the Fermi surface since it gives us the energy gap.  Numerical analysis shows that if momenta $\mathbf{p}$ and $\mathbf{q}$ belong to different valleys, that is, they are located near different Dirac points, we have $A(\mathbf{p};\mathbf{q})\approx0$ and $B(\mathbf{p};\mathbf{q})\approx0$. This property makes it possible to consider the functions $\Delta_{\mathbf{p}\mu}$ near each Dirac point, $\mathbf{K}_1$ and $\mathbf{K}_2$, independently. Let's take for example the $\mathbf{K}_1$ point. The Fermi surface near each Dirac point is a closed curve having near-circular shape. Below we neglect the trigonal warping and approximate the Fermi surface by a circle with radius $q_F^{*}$ calculated numerically by averaging over the Fermi surface. We assume also that $\Delta_{\mathbf{p}\mu}$ are the step-like functions, describing by the equations $\Delta_{\mathbf{K}_1+\mathbf{p}\mu}=\Delta_{\mu}\Theta(q_{\Lambda}-|p-q_F^{*}|)$, where the cutoff momentum $q_{\Lambda}$ will be specified below. Thus, the region of integration in Eq.~\eqref{DeltaP} becomes a ring centered at the Dirac point $\mathbf{K}_1$ and having radii $q_F^{*}-q_{\Lambda}$ and $q_F^{*}+q_{\Lambda}$ if $q_{\Lambda}<q_F^{*}$, or the circle with radius $q_{\Lambda}$, otherwise. In further approximation, we replace the functions $A(\mathbf{p};\mathbf{q})$ and $B(\mathbf{p};\mathbf{q})$ by constants $\bar{A}$ and $\bar{B}$ obtained by averaging of  $A(\mathbf{p};\mathbf{q})$ and $B(\mathbf{p};\mathbf{q})$ over the Fermi surface. We also approximate energies $E_{\mathbf{q}\mu}$ by linear functions
\begin{equation}\label{Eappr1}
E_{\mathbf{K}_1+\mathbf{q}1}\approx v_F^{*}(q-q_F^{*})\,,\;\;E_{\mathbf{K}_1+\mathbf{q}2}\approx v_F^{*}(q+q_F^{*})\,,
\end{equation}
where the renormalized Fermi velocity is calculated numerically by averaging the function $\sum_{\mu}|\partial E_{\mathbf{K}_1+\mathbf{q}\mu}\partial\mathbf{q}|/2$ over the Fermi surface. As a result, the system of equations for order parameters becomes
\begin{eqnarray}
\Delta_{1}\!\!\!&=&\!\!\!\!\int\limits_{q_1}^{q_2}\!\!dq\!\!\left[\frac{\frac12q\lambda_{A}\Delta_{1}}{\sqrt{\Delta_1^2+v_F^{*2}(q-q_F^{*})^2}}\!+\!
\frac{\frac12q\lambda_{B}\Delta_{2}}{\sqrt{\Delta_2^2+v_F^{*2}(q+q_F^{*})^2}}\right]\!,\nonumber\\
\Delta_{2}\!\!\!&=&\!\!\!\!\int\limits_{q_1}^{q_2}\!\!dq\!\!\left[\frac{\frac12q\lambda_{A}\Delta_{2}}{\sqrt{\Delta_2^2+v_F^{*2}(q+q_F^{*})^2}}\!+\!
\frac{\frac12q\lambda_{B}\Delta_{1}}{\sqrt{\Delta_1^2+v_F^{*2}(q-q_F^{*})^2}}\right]\!,\nonumber\\\label{DeltaAppr1}
\end{eqnarray}
where $q_1=q_F^{*}-q_{\Lambda}$, $q_2=q_F^{*}+q_{\Lambda}$ if $q_{\Lambda}<q_F^{*}$, or $q_1=0$, $q_2=q_{\Lambda}$, otherwise, $\lambda_A=2\pi\bar{A}/v_{BZ}$, and $\lambda_B=2\pi\bar{B}/v_{BZ}$. We solve this system numerically.

The magnitude of the order parameters $\Delta_{\mu}$ depends on the values of $\bar{A}$ and $\bar{B}$, as well as on the cut-off momentum $q_{\Lambda}$. Our analysis shows that the main contribution to the functions $A(\mathbf{p};\mathbf{q})$ and $B(\mathbf{p};\mathbf{q})$ comes from the interlayer interaction. Following Refs.~\cite{LozovikSokolik,LozovikPLA2009} we define $q_{\Lambda}$ from the condition $V_{12}(q_{\Lambda})=V_{12}(0)/2$.  Assuming that $q_{\Lambda}d\ll1$ (which is correct for $V_b\lesssim t_0$ and $e^2/\epsilon v_F\lesssim1$), from Eqs.~\eqref{Vrpa} we obtain the estimate $q_{\Lambda}\approx2\pi\Pi_0v_F\alpha$, where $\Pi_0$ is the density of states of the bilayer at the Fermi level, $\alpha=e^2/\epsilon v_F$ is the graphene's fine structure constant, and $v_F=ta\sqrt{3}/2$ is the Fermi velocity of the single layer graphene. Neglecting the trigonal warping, the density of states is expressed as $\Pi_0\approx4q_F^{*}/(\pi v_F^{*})$. Thus, we obtain for $q_{\Lambda}$
\begin{equation}\label{qLambda}
q_{\Lambda}=8\frac{v_F}{v_F^{*}}\alpha q_{F}^{*}\,.
\end{equation}

\textit{Approximate solution, strong interaction limit}---
The limit of weak coupling corresponds the case of $\alpha\ll1$. Simple estimates show, however, that for bilayer suspended in vacuum, $\epsilon=1$, the parameter $\alpha\approx2.6$. When $\alpha$ increases, the cut-off momentum can exceed the size of the superlattice's Brillouin zone. In this case, we should increase the number of bands in our effective Hamiltonian. Simultaneously, we should increase, the number of order parameters $\Delta_{\mathbf{p}\mu}$, with $\mu$ now changing from $1$ to some $N>2$. The rank of the matrix functions $A_{\mu\nu}$ and $B_{\mu\nu}$ becomes equal to $N$. This consideration can be substantially simplified if we consider  the added `high-energy' bands in the limit of decoupled ($t_0=0$) graphene layers. This approximation is justified, since for these bands we have $|E_\mathbf{p}^{(S)}|\gtrsim t_0$. For decoupled layers, one can associate the band index $S$ to the momentum $\mathbf{p}$ lying in the Brillouin zone of the layer $1$ or $2$, that is, one can perform the band unfolding~\cite{BandUnfoldingPRB2017}. As a result, one can assume that the number of order parameters $\Delta_{\mathbf{p}\mu}$ is still equal to $2$, but the integration in Eqs.~\eqref{DeltaPSigma} or~\eqref{DeltaP} is extended to the momenta exceeding the reciprocal  unit cell of the superlattice. Applying this procedure, one must keep in mind that the two valleys should be still considered independently. To understand why this is so, let us consider the situation in the unfolded Brillouin zone from the beginning. Layer $1$ has two non-equivalent Dirac points, $\mathbf{K}$ and $\mathbf{K}'$($=-\mathbf{K}$). Rotation by twist angle $\theta$ transforms them into Dirac points of the layer $2$, $\mathbf{K}_{\theta}$ and $\mathbf{K}'_{\theta}$. The considered ordering corresponds to the formation of the electron-hole pair consisting of the electron with momentum $\mathbf{K}+\mathbf{p}$ and the hole with momentum $\mathbf{K}_{\theta}'+\mathbf{p}$ (valley $1$), and the electron with momentum $\mathbf{K}'+\mathbf{p}$ and the hole with momentum $\mathbf{K}_{\theta}+\mathbf{p}$ (valley $2$).~\footnote{For superstructures with $r\neq3n$. Otherwise, the electron and the hole have momenta $\mathbf{K}+\mathbf{p}$ and $\mathbf{K}_{\theta}+\mathbf{p}$ in valley 1, and $\mathbf{K}'+\mathbf{p}$ and $\mathbf{K}_{\theta}'+\mathbf{p}$ in valley 2.} It is clear, that the annihilation of such a pair in the valley $1$ with simultaneous creation of the pair in the valley $2$ is prohibited by the momentum conservation law. Thus, in our model, the intervalley scattering can be neglected~\footnote{This picture is valid, of course, for $E_{\Lambda}\equiv v_Fq_{\Lambda}\lesssim t$. At larger energies, the spectrum cannot be described by separate Dirac cones.}.

The procedure described above implies the knowledge of the energies $E_{\mathbf{p}\mu}$ at large momenta. We approximate $E_{\mathbf{K}_1+\mathbf{p}\mu}$ by Eqs.~\eqref{Eappr1}, when $p<q_0$, where $q_0=|\bm{{\cal G}}_{1,2}|/(2\sqrt{3})$ is the radius of the circle centered at Dirac point $\mathbf{K}_1$ and touching the edges of the reciprocal  unit cell of the superlattice (see Fig.~\ref{FigSpec}).  At larger $p$ we use the limit of decoupled layers
\begin{equation}\label{Eappr2}
E_{\mathbf{K}_1+\mathbf{p}1}\approx E_{\mathbf{K}_1+\mathbf{p}2}\approx v_Fp\,.
\end{equation}
With this accuracy we neglect the effect of the bias voltage at high energies, since $V_b\lesssim t_0$. As a result, approximate equation for the order parameters $\Delta_{\mu}$ become
\begin{eqnarray}
\Delta_{1}\!\!\!&=&\!\!\!\!\int\limits_{0}^{q_0}\!\!dq\frac{\frac12q\lambda_{A}\Delta_{1}}{\sqrt{\Delta_1^2+v_F^{*2}(q-q_F^{*})^2}}+
\int\limits_{q_0}^{q_{\Lambda}}\!\!dq\frac{\frac12q\lambda_{A}\Delta_{1}}{\sqrt{\Delta_1^2+v_F^{2}q^2}}+\nonumber\\
&&\!\!\!\!\int\limits_{0}^{q_0}\!\!dq\frac{\frac12q\lambda_{B}\Delta_{2}}{\sqrt{\Delta_2^2+v_F^{*2}(q+q_F^{*})^2}}+
\int\limits_{q_0}^{q_{\Lambda}}\!\!dq\frac{\frac12q\lambda_{B}\Delta_{2}}{\sqrt{\Delta_2^2+v_F^{2}q^2}}\,,\nonumber\\
\Delta_{2}\!\!\!&=&\!\!\!\!\int\limits_{0}^{q_0}\!\!dq\frac{\frac12q\lambda_{A}\Delta_{2}}{\sqrt{\Delta_2^2+v_F^{*2}(q+q_F^{*})^2}}+
\int\limits_{q_0}^{q_{\Lambda}}\!\!dq\frac{\frac12q\lambda_{A}\Delta_{2}}{\sqrt{\Delta_2^2+v_F^{2}q^2}}+\nonumber\\
&&\!\!\!\!\int\limits_{0}^{q_0}\!\!dq\frac{\frac12q\lambda_{B}\Delta_{1}}{\sqrt{\Delta_1^2+v_F^{*2}(q-q_F^{*})^2}}+
\int\limits_{q_0}^{q_{\Lambda}}\!\!dq\frac{\frac12q\lambda_{B}\Delta_{1}}{\sqrt{\Delta_1^2+v_F^{2}q^2}}\,.\nonumber\\\label{DeltaAppr2}
\end{eqnarray}

\begin{figure}[t]
\centering
\includegraphics[width=0.98\columnwidth]{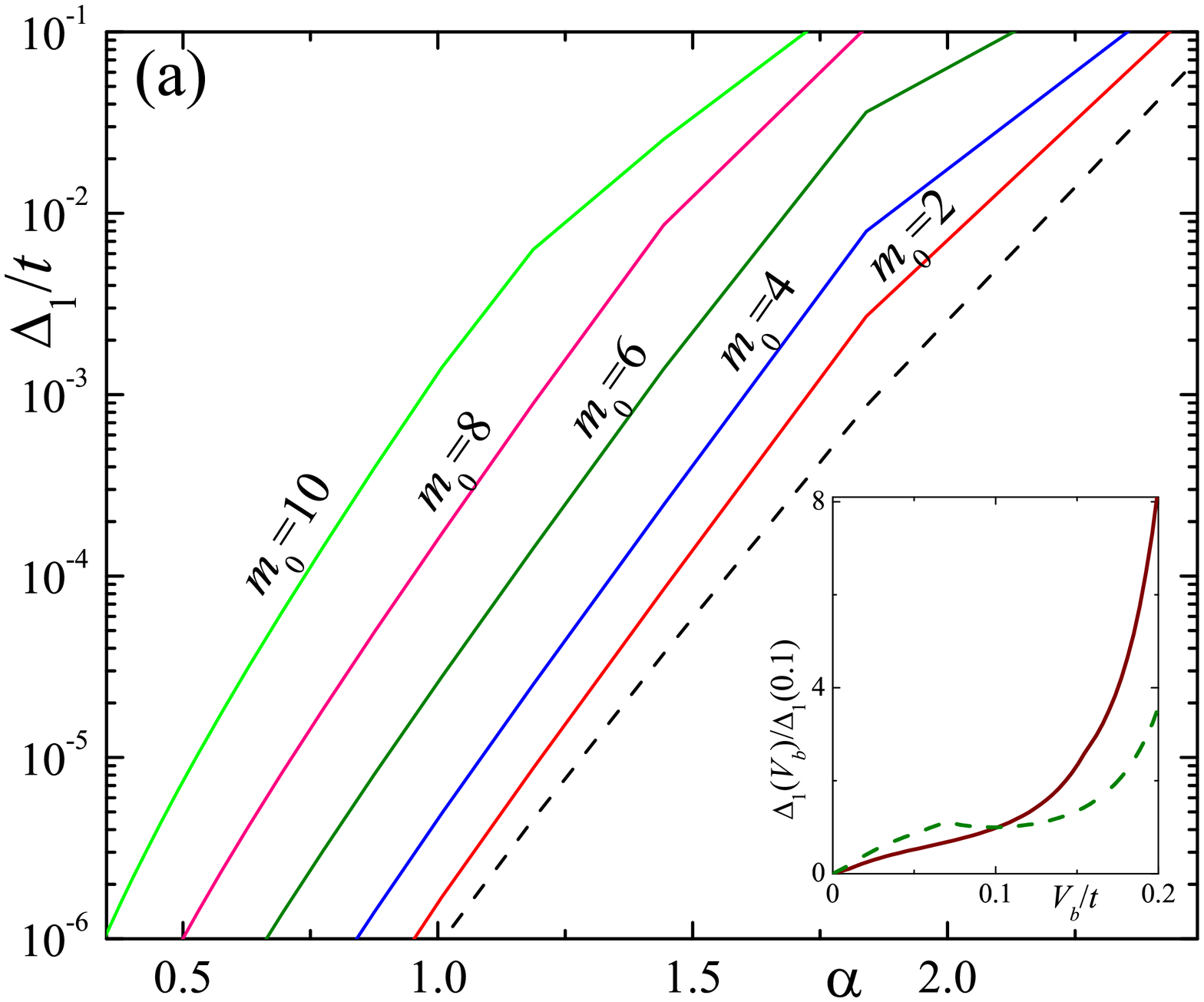}\\
\vspace{2mm}
\includegraphics[width=0.98\columnwidth]{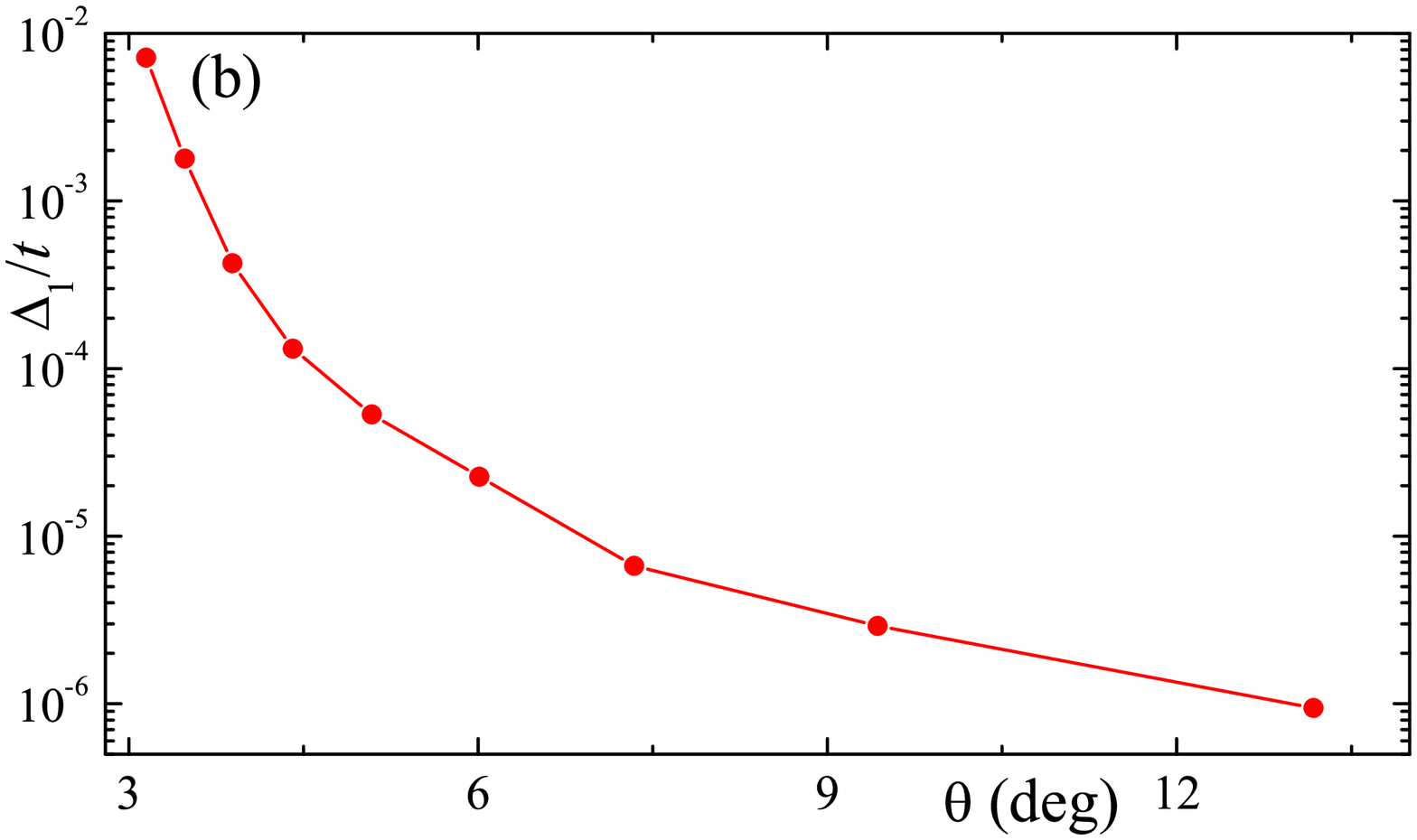}
\caption{(a) Solid curves show the dependence of $\Delta_1$ on $\alpha$ for superstructures with $r=1$ and $m_0=2,\,4,\,6,\,8,\,10$ (twist angles $\theta\cong17.7,\,7.3,\,5.1,\,3.9,\,3.2$ degrees) provided that $V_b/t=0.1$. Dashed curve corresponds to the hypothetical case of decoupled ($t_0=0$) graphene layers. In this case, the function $\Delta_1(\alpha)$ is almost independent on $\theta$. Inset shows the dependencies of $\Delta_1$ on $V_b$ for $\alpha=0.306$ (solid curve) and for $\alpha=0.719$ (dashed curve) calculated for superstructure $m_0=5$, $r=1$. (b) The dependence of $\Delta_1$ on $\theta$ calculated for $V_b/t=0.037$ and $\alpha=1.044$.\label{FigDelta}}
\end{figure}

\textit{Results and Discussion}---
We solve the system of equations~\eqref{DeltaAppr2} [or~\eqref{DeltaAppr1}, if $q_{\Lambda}<q_0$] numerically for several superstructures with $r=1$ in a wide range of $V_b$ and $\alpha$. We found that the ratio $\Delta_{2}/\Delta_{1}\gtrsim0.3$ for any interaction strength considered and goes to $1$ when $\alpha$ increases. Our main interest is the value of $\Delta_1$ since it gives us the energy gap. The dependencies of the band gap $\Delta_1$ on $\alpha$ for superstructures with $m_0=2,\,4,\,6,\,8,\,10$ (corresponding twist angles are $\theta\cong17.7,\,7.3,\,5.1,\,3.9,\,3.2$ degrees) are presented in Fig.~\ref{FigDelta}(a). The gap strongly (exponentially) depends on the interaction strength for all superstructures. It is seen from this figure that the gap is considerably large only when $\alpha\gtrsim1$, that is, when $\epsilon\lesssim2.5$. The ratio $\Delta_1/t\sim10^{-2}$ corresponds to the value of the gap about $300$\,K. Thus, according to our calculations, in order to observe the gap at room temperatures, the permittivity of the sample should not be large.

The important result demonstrating in Fig.~\ref{FigDelta}(a) is that for any $\alpha$, the band gap is larger for superstructures with smaller twist angles (larger $m_0$). This is illustrated in Fig.~\ref{FigDelta}(b), where we plot the dependence of $\Delta_1$ on $\theta$ calculated at $\alpha=1.044$ and $V_b/t=0.037$. We see that the band gap increases by about $4$ orders of magnitude, when twist angle changes from $\theta\cong17.7^{\circ}$ ($m_0=2$) down to $\theta\cong3.2^{\circ}$ ($m_0=10$). Such a strong enhancement can be explained by the reduction of the Fermi velocity due to the interlayer hybridization. To illustrate this, let us consider the weak interaction limit. Assuming that $q_{\lambda}\ll q_F^{*}$ and $\Delta_{\mu}\ll V_b$, one can solve the system of equations~\eqref{DeltaAppr1} analytically. This gives
\begin{equation}\label{DeltaApprAn}
\Delta_1\approx2v_Fq_F^{*}\alpha e^{-1/\Lambda+4\alpha^{*}},
\end{equation}
where $\alpha^{*}=e^2/(\epsilon v_F^{*})$ is the `renormalized $\alpha$' and
\begin{equation}\label{Lambda}
\Lambda\approx\frac{2\pi\bar{A}q_F^{*}}{v_{BZ}v_F^{*}}\,.
\end{equation}
Calculations show that good approximation for $\bar{A}$ and, consequently, for $\Lambda$ can be obtained if in Eq.~\eqref{AB} for $A_{\mu\nu}(\mathbf{p};\mathbf{q})$
we will use wave functions corresponding to the limit of decoupled layers, $t_0=0$. In this case, expressions for $U^{(S)}_{\mathbf{pG}i\alpha}$ can be found analytically, and as a result, we obtain
\begin{equation}\label{AAppr}
\bar{A}\approx\left\langle\cos^2(\varphi/2)V_{12}\left(2q_F^{*}|\sin(\varphi/2)|\right)\right\rangle\,,
\end{equation}
where $\varphi$ is the polar angle parameterizing points on the Fermi surface, while averaging is performed over this angle. The factor $\cos^2(\varphi/2)$ before potential $V_{12}$ is inherited from the wave functions. Substituting this expression into Eq.~\eqref{Lambda}, we obtain
\begin{equation}\label{LambdaAppr}
\Lambda\approx\frac18\left\langle\frac{\cos^2(\varphi/2)}{1+\displaystyle\frac{1}{4\alpha^{*}}|\sin(\varphi/2)|}\right\rangle.
\end{equation}
Deriving the latter equality we assumed that $q_F^{*}d\ll1$, which is valid for all bias voltages considered. According to Eq.~\eqref{LambdaAppr}, $\Lambda$  increases when $\alpha^{*}$ increases. When $\alpha^{*}\ll1$, we have $\Lambda\propto\alpha^{*}\ln(1/\alpha^{*})$. Since $\alpha^{*}$ inversely proportional to $v_F^{*}$, parameter $\Lambda$ and, consequently, $\Delta_1$ increases when $v_F^{*}$ decreases. This can explain the dependence of $\Delta_1$ on $\theta$ since $v_F^{*}$ decreases with the twist angle~\cite{dSPRL,PNAS,ourTBLG}. Numerical calculations show also, that actual value of $\bar{A}$ is greater than estimate~\eqref{AAppr}. This can be explained by the fact that at finite interlayer hybridization, the quasiparticles are no longer localized in one particular layer. As a result, intralayer potential contributes to $\bar{A}$, and this effect is stronger for smaller twist angles.

It is seen from Eqs.~\eqref{DeltaApprAn}, that exponent in Eq.~\eqref{LambdaAppr} is independent on the bias voltage. This result correlates with that obtained in Refs.~\cite{LozovikSokolik,LozovikPLA2009}. Such a behavior can be understood if we realize that parameter $\Lambda$ is a product of the interaction strength $\bar{A}\propto1/q_F^{*}$ and the density of states at the Fermi level $\rho_0\propto q_F^{*}\propto V_b$. Thus, only pre-exponential factor depends the bias voltage. Numerical analysis shows, that at small $V_b$, the band gap linearly depends on $V_b$ for any $\alpha$, while at strong interaction the function $\Delta_1(V_b)$ can show non-monotonous behavior at larger bias voltages [see the inset to Fig.~\ref{FigDelta}(a)].

The order parameter goes to zero when $V_b\to0$ for any superstructures and for any strength of the interaction considered. At the same time, it is seen from Fig.~\ref{FigDelta}(b), that $\Delta_1$ increases very fast with the decrease of the twist angle. We did not analyze what happens at twist angles very close and below critical value ($\theta_c\cong1.89^{\circ}$ in our model), but we can expect that for these $\theta$s the Coulomb interaction can stabilize some type of ordering even at zero bias voltage. This suggestion is confirmed by recent studies performed in Ref.~\cite{GuineaArXiv2017}, where authors predict the SDW ground state for bilayers with $\theta<\theta_c$ in the framework of the Hubbard model. Note that our `exciton-plus-SDW' order parameter is stabilized mainly by the interlayer interaction, and it would not occur (or would be strongly suppressed) for Hubbard interaction. Similar type of order has been considered in Ref.~\cite{AkzyanovAABLG2014} devoted to the study of the AA-stacked bilayer graphene in the applied electric field. The detailed investigation of the ordering type for bilayers with the twist angles close to the critical value is the subject for future study.

In conclusion, we studied the ground state of the twisted bilayer graphene when the bias voltage is applied to the system. We showed that the bias voltage forms two hole-like and two electron-like Fermi surfaces with perfect nesting. As a result of such a band structure, the screened Coulomb interaction stabilizes the exciton order parameter in the system. The exciton order parameter is accompanied by the spin-density-wave order. The value of the gap depends on the twist angle and on the applied voltage. The latter property is quite useful for different applications in electronics.

\textit{Acknowledgments.}--- This work is partially supported by the
Russian Foundation for Basic Research (Projects 17-02-00323, and
15-02-02128), JSPS-RFBR grant No 17-52-50023,
the RIKEN iTHES Project,
MURI Center for Dynamic Magneto-Optics via the AFOSR Award
No.~FA9550-14-1-0040,
the Japan Society for the Promotion of Science (KAKENHI),
the IMPACT program of JST,
JSPS-RFBR grant No 17-52-50023,
CREST grant No.~JPMJCR1676,
and the Sir John Templeton Foundation.

\bibliographystyle{apsrevlong_no_issn_url}

\begin{thebibliography}{19}
\expandafter\ifx\csname natexlab\endcsname\relax\def\natexlab#1{#1}\fi
\expandafter\ifx\csname bibnamefont\endcsname\relax
  \def\bibnamefont#1{#1}\fi
\expandafter\ifx\csname bibfnamefont\endcsname\relax
  \def\bibfnamefont#1{#1}\fi
\expandafter\ifx\csname citenamefont\endcsname\relax
  \def\citenamefont#1{#1}\fi

\bibitem[{\citenamefont{Rozhkov et~al.}(2016)\citenamefont{Rozhkov, Sboychakov,
  Rakhmanov, and Nori}}]{ourBLGreview2016}
\bibinfo{author}{\bibfnamefont{A.}~\bibnamefont{Rozhkov}},
  \bibinfo{author}{\bibfnamefont{A.}~\bibnamefont{Sboychakov}},
  \bibinfo{author}{\bibfnamefont{A.}~\bibnamefont{Rakhmanov}},
  \bibnamefont{and} \bibinfo{author}{\bibfnamefont{F.}~\bibnamefont{Nori}},
  {``}\bibinfo{title}{Electronic properties of graphene-based bilayer
  systems},{''} \bibinfo{journal}{Physics Reports}
  \textbf{\bibinfo{volume}{648}}, \bibinfo{pages}{1} (\bibinfo{year}{2016}).

\bibitem[{\citenamefont{Mele}(2012)}]{MeleReview}
\bibinfo{author}{\bibfnamefont{E.~J.} \bibnamefont{Mele}},
  {``}\bibinfo{title}{Interlayer coupling in rotationally faulted multilayer
  graphenes},{''} \bibinfo{journal}{Journal of Physics D: Applied Physics}
  \textbf{\bibinfo{volume}{45}}, \bibinfo{pages}{154004}
  (\bibinfo{year}{2012}).

\bibitem[{\citenamefont{Brihuega et~al.}(2012)\citenamefont{Brihuega, Mallet,
  Gonz\'alez-Herrero, Trambly~de Laissardi\`ere, Ugeda, Magaud,
  G\'omez-Rodr\'{\i}guez, Yndur\'ain, and Veuillen}}]{STM1}
\bibinfo{author}{\bibfnamefont{I.}~\bibnamefont{Brihuega}},
  \bibinfo{author}{\bibfnamefont{P.}~\bibnamefont{Mallet}},
  \bibinfo{author}{\bibfnamefont{H.}~\bibnamefont{Gonz\'alez-Herrero}},
  \bibinfo{author}{\bibfnamefont{G.}~\bibnamefont{Trambly~de Laissardi\`ere}},
  \bibinfo{author}{\bibfnamefont{M.~M.} \bibnamefont{Ugeda}},
  \bibinfo{author}{\bibfnamefont{L.}~\bibnamefont{Magaud}},
  \bibinfo{author}{\bibfnamefont{J.~M.} \bibnamefont{G\'omez-Rodr\'{\i}guez}},
  \bibinfo{author}{\bibfnamefont{F.}~\bibnamefont{Yndur\'ain}},
  \bibnamefont{and} \bibinfo{author}{\bibfnamefont{J.-Y.}
  \bibnamefont{Veuillen}}, {``}\bibinfo{title}{Unraveling the Intrinsic and
  Robust Nature of van Hove Singularities in Twisted Bilayer Graphene by
  Scanning Tunneling Microscopy and Theoretical Analysis},{''}
  \bibinfo{journal}{Phys. Rev. Lett.} \textbf{\bibinfo{volume}{109}},
  \bibinfo{pages}{196802} (\bibinfo{year}{2012}).

\bibitem[{\citenamefont{Luican et~al.}(2011)\citenamefont{Luican, Li, Reina,
  Kong, Nair, Novoselov, Geim, and Andrei}}]{STM2}
\bibinfo{author}{\bibfnamefont{A.}~\bibnamefont{Luican}},
  \bibinfo{author}{\bibfnamefont{G.}~\bibnamefont{Li}},
  \bibinfo{author}{\bibfnamefont{A.}~\bibnamefont{Reina}},
  \bibinfo{author}{\bibfnamefont{J.}~\bibnamefont{Kong}},
  \bibinfo{author}{\bibfnamefont{R.~R.} \bibnamefont{Nair}},
  \bibinfo{author}{\bibfnamefont{K.~S.} \bibnamefont{Novoselov}},
  \bibinfo{author}{\bibfnamefont{A.~K.} \bibnamefont{Geim}}, \bibnamefont{and}
  \bibinfo{author}{\bibfnamefont{E.~Y.} \bibnamefont{Andrei}},
  {``}\bibinfo{title}{Single-Layer Behavior and Its Breakdown in Twisted
  Graphene Layers},{''} \bibinfo{journal}{Phys. Rev. Lett.}
  \textbf{\bibinfo{volume}{106}}, \bibinfo{pages}{126802}
  (\bibinfo{year}{2011}).

\bibitem[{\citenamefont{Ohta et~al.}(2012)\citenamefont{Ohta, Beechem,
  Robinson, and Kellogg}}]{SynthesysMoire}
\bibinfo{author}{\bibfnamefont{T.}~\bibnamefont{Ohta}},
  \bibinfo{author}{\bibfnamefont{T.~E.} \bibnamefont{Beechem}},
  \bibinfo{author}{\bibfnamefont{J.~T.} \bibnamefont{Robinson}},
  \bibnamefont{and} \bibinfo{author}{\bibfnamefont{G.~L.}
  \bibnamefont{Kellogg}}, {``}\bibinfo{title}{Long-range atomic ordering and
  variable interlayer interactions in two overlapping graphene lattices with
  stacking misorientations},{''} \bibinfo{journal}{Phys. Rev. B}
  \textbf{\bibinfo{volume}{85}}, \bibinfo{pages}{075415}
  (\bibinfo{year}{2012}).

\bibitem[{\citenamefont{Lopes~dos Santos et~al.}(2012)\citenamefont{Lopes~dos
  Santos, Peres, and Castro~Neto}}]{dSPRB}
\bibinfo{author}{\bibfnamefont{J.~M.~B.} \bibnamefont{Lopes~dos Santos}},
  \bibinfo{author}{\bibfnamefont{N.~M.~R.} \bibnamefont{Peres}},
  \bibnamefont{and} \bibinfo{author}{\bibfnamefont{A.~H.}
  \bibnamefont{Castro~Neto}}, {``}\bibinfo{title}{Continuum model of the
  twisted graphene bilayer},{''} \bibinfo{journal}{Phys. Rev. B}
  \textbf{\bibinfo{volume}{86}}, \bibinfo{pages}{155449}
  (\bibinfo{year}{2012}).

\bibitem[{\citenamefont{Mele}(2010)}]{MelePRB1}
\bibinfo{author}{\bibfnamefont{E.~J.} \bibnamefont{Mele}},
  {``}\bibinfo{title}{Commensuration and interlayer coherence in twisted
  bilayer graphene},{''} \bibinfo{journal}{Phys. Rev. B}
  \textbf{\bibinfo{volume}{81}}, \bibinfo{pages}{161405}
  (\bibinfo{year}{2010}).

\bibitem[{\citenamefont{Shallcross et~al.}(2010)\citenamefont{Shallcross,
  Sharma, Kandelaki, and Pankratov}}]{Pankratov1}
\bibinfo{author}{\bibfnamefont{S.}~\bibnamefont{Shallcross}},
  \bibinfo{author}{\bibfnamefont{S.}~\bibnamefont{Sharma}},
  \bibinfo{author}{\bibfnamefont{E.}~\bibnamefont{Kandelaki}},
  \bibnamefont{and} \bibinfo{author}{\bibfnamefont{O.~A.}
  \bibnamefont{Pankratov}}, {``}\bibinfo{title}{Electronic structure of
  turbostratic graphene},{''} \bibinfo{journal}{Phys. Rev. B}
  \textbf{\bibinfo{volume}{81}}, \bibinfo{pages}{165105}
  (\bibinfo{year}{2010}).

\bibitem[{\citenamefont{Sboychakov et~al.}(2015)\citenamefont{Sboychakov,
  Rakhmanov, Rozhkov, and Nori}}]{ourTBLG}
\bibinfo{author}{\bibfnamefont{A.~O.} \bibnamefont{Sboychakov}},
  \bibinfo{author}{\bibfnamefont{A.~L.} \bibnamefont{Rakhmanov}},
  \bibinfo{author}{\bibfnamefont{A.~V.} \bibnamefont{Rozhkov}},
  \bibnamefont{and} \bibinfo{author}{\bibfnamefont{F.}~\bibnamefont{Nori}},
  {``}\bibinfo{title}{Electronic spectrum of twisted bilayer graphene},{''}
  \bibinfo{journal}{Phys. Rev. B} \textbf{\bibinfo{volume}{92}},
  \bibinfo{pages}{075402} (\bibinfo{year}{2015}).

\bibitem[{\citenamefont{Rozhkov et~al.}(2017)\citenamefont{Rozhkov, Sboychakov,
  Rakhmanov, and Nori}}]{ourTBLG2017}
\bibinfo{author}{\bibfnamefont{A.}~\bibnamefont{Rozhkov}},
  \bibinfo{author}{\bibfnamefont{A.}~\bibnamefont{Sboychakov}},
  \bibinfo{author}{\bibfnamefont{A.}~\bibnamefont{Rakhmanov}},
  \bibnamefont{and} \bibinfo{author}{\bibfnamefont{F.}~\bibnamefont{Nori}},
  {``}\bibinfo{title}{Single-electron gap in the spectrum of twisted bilayer
  graphene},{''} \bibinfo{journal}{Physical Review B}
  \textbf{\bibinfo{volume}{95}}, \bibinfo{pages}{045119}
  (\bibinfo{year}{2017}).

\bibitem[{\citenamefont{Lopes~dos Santos et~al.}(2007)\citenamefont{Lopes~dos
  Santos, Peres, and Castro~Neto}}]{dSPRL}
\bibinfo{author}{\bibfnamefont{J.~M.~B.} \bibnamefont{Lopes~dos Santos}},
  \bibinfo{author}{\bibfnamefont{N.~M.~R.} \bibnamefont{Peres}},
  \bibnamefont{and} \bibinfo{author}{\bibfnamefont{A.~H.}
  \bibnamefont{Castro~Neto}}, {``}\bibinfo{title}{Graphene Bilayer with a
  Twist: Electronic Structure},{''} \bibinfo{journal}{Phys. Rev. Lett.}
  \textbf{\bibinfo{volume}{99}}, \bibinfo{pages}{256802}
  (\bibinfo{year}{2007}).

\bibitem[{\citenamefont{Bistritzer and MacDonald}(2011)}]{PNAS}
\bibinfo{author}{\bibfnamefont{R.}~\bibnamefont{Bistritzer}} \bibnamefont{and}
  \bibinfo{author}{\bibfnamefont{A.~H.} \bibnamefont{MacDonald}},
  {``}\bibinfo{title}{Moir\'{e} bands in twisted double-layer graphene},{''}
  \bibinfo{journal}{Proceedings of the National Academy of Sciences}
  \textbf{\bibinfo{volume}{108}}, \bibinfo{pages}{12233}
  (\bibinfo{year}{2011}).

\bibitem[{\citenamefont{Su\'arez~Morell
  et~al.}(2010)\citenamefont{Su\'arez~Morell, Correa, Vargas, Pacheco, and
  Barticevic}}]{Morell1}
\bibinfo{author}{\bibfnamefont{E.}~\bibnamefont{Su\'arez~Morell}},
  \bibinfo{author}{\bibfnamefont{J.~D.} \bibnamefont{Correa}},
  \bibinfo{author}{\bibfnamefont{P.}~\bibnamefont{Vargas}},
  \bibinfo{author}{\bibfnamefont{M.}~\bibnamefont{Pacheco}}, \bibnamefont{and}
  \bibinfo{author}{\bibfnamefont{Z.}~\bibnamefont{Barticevic}},
  {``}\bibinfo{title}{Flat bands in slightly twisted bilayer graphene:
  Tight-binding calculations},{''} \bibinfo{journal}{Phys. Rev. B}
  \textbf{\bibinfo{volume}{82}}, \bibinfo{pages}{121407}
  (\bibinfo{year}{2010}).

\bibitem[{\citenamefont{Trambly~de Laissardi\`{e}re
  et~al.}(2010)\citenamefont{Trambly~de Laissardi\`{e}re, Mayou, and
  Magaud}}]{NanoLettTB}
\bibinfo{author}{\bibfnamefont{G.}~\bibnamefont{Trambly~de Laissardi\`{e}re}},
  \bibinfo{author}{\bibfnamefont{D.}~\bibnamefont{Mayou}}, \bibnamefont{and}
  \bibinfo{author}{\bibfnamefont{L.}~\bibnamefont{Magaud}},
  {``}\bibinfo{title}{Localization of Dirac Electrons in Rotated Graphene
  Bilayers},{''} \bibinfo{journal}{Nano Letters} \textbf{\bibinfo{volume}{10}},
  \bibinfo{pages}{804} (\bibinfo{year}{2010}).

\bibitem[{\citenamefont{Lozovik and Sokolik}(2008)}]{LozovikSokolik}
\bibinfo{author}{\bibfnamefont{Y.~E.} \bibnamefont{Lozovik}} \bibnamefont{and}
  \bibinfo{author}{\bibfnamefont{A.}~\bibnamefont{Sokolik}},
  {``}\bibinfo{title}{Electron-hole pair condensation in a graphene
  bilayer},{''} \bibinfo{journal}{JETP letters} \textbf{\bibinfo{volume}{87}},
  \bibinfo{pages}{55} (\bibinfo{year}{2008}).

\bibitem[{\citenamefont{Lozovik and Sokolik}(2009)}]{LozovikPLA2009}
\bibinfo{author}{\bibfnamefont{Y.~E.} \bibnamefont{Lozovik}} \bibnamefont{and}
  \bibinfo{author}{\bibfnamefont{A.}~\bibnamefont{Sokolik}},
  {``}\bibinfo{title}{Multi-band pairing of ultrarelativistic electrons and
  holes in graphene bilayer},{''} \bibinfo{journal}{Physics Letters A}
  \textbf{\bibinfo{volume}{374}}, \bibinfo{pages}{326} (\bibinfo{year}{2009}).

\bibitem[{\citenamefont{Nishi et~al.}(2017)\citenamefont{Nishi, Matsushita, and
  Oshiyama}}]{BandUnfoldingPRB2017}
\bibinfo{author}{\bibfnamefont{H.}~\bibnamefont{Nishi}},
  \bibinfo{author}{\bibfnamefont{Y.-i.} \bibnamefont{Matsushita}},
  \bibnamefont{and} \bibinfo{author}{\bibfnamefont{A.}~\bibnamefont{Oshiyama}},
  {``}\bibinfo{title}{Band-unfolding approach to moir\'e-induced band-gap
  opening and Fermi level velocity reduction in twisted bilayer graphene},{''}
  \bibinfo{journal}{Phys. Rev. B} \textbf{\bibinfo{volume}{95}},
  \bibinfo{pages}{085420} (\bibinfo{year}{2017}).

\bibitem[{\citenamefont{Gonzalez-Arraga
  et~al.}(2017)\citenamefont{Gonzalez-Arraga, Lado, Guine, and
  San-Jose}}]{GuineaArXiv2017}
\bibinfo{author}{\bibfnamefont{L.~A.} \bibnamefont{Gonzalez-Arraga}},
  \bibinfo{author}{\bibfnamefont{J.}~\bibnamefont{Lado}},
  \bibinfo{author}{\bibfnamefont{F.}~\bibnamefont{Guine}}, \bibnamefont{and}
  \bibinfo{author}{\bibfnamefont{P.}~\bibnamefont{San-Jose}},
  {``}\bibinfo{title}{Electrically controllable magnetism in twisted bilayer
  graphene},{''} \bibinfo{journal}{arXiv preprint arXiv:1702.08831}
  (\bibinfo{year}{2017}).

\bibitem[{\citenamefont{Akzyanov et~al.}(2014)\citenamefont{Akzyanov,
  Sboychakov, Rozhkov, Rakhmanov, and Nori}}]{AkzyanovAABLG2014}
\bibinfo{author}{\bibfnamefont{R.~S.} \bibnamefont{Akzyanov}},
  \bibinfo{author}{\bibfnamefont{A.~O.} \bibnamefont{Sboychakov}},
  \bibinfo{author}{\bibfnamefont{A.~V.} \bibnamefont{Rozhkov}},
  \bibinfo{author}{\bibfnamefont{A.~L.} \bibnamefont{Rakhmanov}},
  \bibnamefont{and} \bibinfo{author}{\bibfnamefont{F.}~\bibnamefont{Nori}},
  {``}\bibinfo{title}{$AA$-stacked bilayer graphene in an applied electric
  field: Tunable antiferromagnetism and coexisting exciton order
  parameter},{''} \bibinfo{journal}{Phys. Rev. B}
  \textbf{\bibinfo{volume}{90}}, \bibinfo{pages}{155415}
  (\bibinfo{year}{2014}).

\end{thebibliography}

\end{document}